\documentclass[aps,prb,twocolumn,amsmath,amssymb,showpacs,groupedaddress]{revtex4-1}

\usepackage{graphicx}
\usepackage{hyperref}
\usepackage{color}

\makeatletter
\@ifundefined{textcolor}{}
{
 \definecolor{BLACK}{gray}{0}
 \definecolor{WHITE}{gray}{1}
 \definecolor{RED}{rgb}{1,0,0}
 \definecolor{GREEN}{rgb}{0,1,0}
 \definecolor{BLUE}{rgb}{0,0,1}
 \definecolor{CYAN}{cmyk}{1,0,0,0}
 \definecolor{MAGENTA}{cmyk}{0,1,0,0}
 \definecolor{YELLOW}{cmyk}{0,0,1,0}
}

\begin{document}

\title{Three-dimensionality of the bulk electronic structure in WTe${}_{2}$}

\author{Yun Wu}

\author{Na Hyun Jo}

\author{Daixiang Mou}

\author{Lunan Huang}

\author{S.~L.~Bud'ko}

\author{P. C. Canfield}
\email[]{canfield@ameslab.gov}

\author{Adam Kaminski}
\email[]{kaminski@ameslab.gov}
\affiliation{ Ames Laboratory, U.S. DOE and Department of Physics and Astronomy, Iowa State University, Ames, Iowa 50011, USA}

\date{\today}

\begin{abstract}
We use temperature- and field-dependent resistivity measurements [Shubnikov--de Haas (SdH) quantum oscillations] and ultrahigh resolution, tunable, vacuum ultraviolet (VUV) laser-based angle-resolved photoemission spectroscopy (ARPES) to study the three-dimensionality (3D) of the bulk electronic structure in WTe${}_{2}$, a type-II Weyl semimetal. The bulk Fermi surface (FS) consists of two pairs of electron pockets and two pairs of hole pockets along the $X-\Gamma-X$ direction as detected by using an incident photon energy of 6.7~eV, which is consistent with the previously reported data. However, if using an incident photon energy of 6.36~eV, another pair of tiny electron pockets is detected on both sides of the $\Gamma$ point, which is in agreement with the small quantum oscillation frequency peak observed in the magnetoresistance. Therefore, the bulk, 3D FS consists of three pairs of electron pockets and two pairs of hole pockets in total. With the ability of fine tuning the incident photon energy, we demonstrate the strong three-dimensionality of the bulk electronic structure in WTe${}_{2}$. The combination of resistivity and ARPES measurements reveal the complete, and consistent, picture of the bulk electronic structure of this material. 
\end{abstract}

\maketitle

\section{Introduction}

Extremely large magnetoresistance, i.e. dramatic increase in the resistance of a material upon applied magnetic fields, has recently attracted great interest~\cite{Mun12PRB, Ali14Nat, Liang15NatMat}. Materials with this type of property can be potentially very useful for applications such as magnetic field sensors, data storage and processing etc. Interestingly, some of these extremely large magnetoresistive materials are hosts for other exotic properties. Among the first few extremely large magnetoresistive materials, PtSn${}_{4}$~\cite{Mun12PRB} has been reported to host unusual Dirac node arc structure, that is the Dirac dispersion extending in momentum space in one dimension and gapped out at both ends~\cite{Wu2016Dirac}. Another material with extremely large magnetoresistance, Cd${}_{3}$As${}_{2}$~\cite{Liang15NatMat} was shown to be one of the first three-dimensional Dirac semimetals with linear dispersion along all three momentum directions~\cite{Wang13PRB, Neupane14NatCom, Liu14NatMat, Borisenko14PRL, Narayanan15PRL}. Extremely large magnetoresistive material, WTe${}_{2}$~\cite{Ali14Nat} has been reported to exhibit pressure-induced superconductivity~\cite{Kang15NatComm, Pan15NatComm}, and a pressure-induced Lifshitz phase transition was proposed to explain the emergence of the superconductivity~\cite{Kang15NatComm}. Surprisingly, a temperature-induced Lifshitz transition was recently reported in WTe${}_{2}$. The significant shift of the chemical potential with moderate temperature change is caused by the close proximity of electron and hole band extrema to the chemical potential~\cite{Wu15PRL}. More interestingly, WTe${}_{2}$ was the first material proposed to be a type-II Weyl semimetal~\cite{Soluyanov2015Type}. Unlike the type-I Weyl semimetals~\cite{Huang15NatCom, Weng15PRX, Xu15SciDis, Yang15NatPhys, Lv15NatPhys, Lv2015Experimental, Xu15NatPhys}, such materials have the Weyl points emerging at the touching points of the electron and hole pockets~\cite{Soluyanov2015Type}. Recently, multiple ARPES measurements reported the presence of the Fermi arc surface states in these compounds~\cite{Belopolski2016Fermi, Huang2016Spectroscopic, Tamai2016Fermi, Deng2016Experimental, Jiang2016Observation, Liang2016Electronic, Xu2016Discovery, Bruno2016Surface, Wang2016Observation, Wu2016Observation, Feng2016Spin}. Photon energy dependence measurements have been used to demonstrate the two-dimensionality (surface origin) of the Fermi arc in WTe${}_{2}$~\cite{Bruno2016Surface, Wu2016Observation}. However, detailed measurements of three-dimensional bulk electronic structures are still lacking. 

ARPES has been known as the most direct technique for probing the electronic structures of materials~\cite{Damascelli03RMP, Campuzano04PS}. Early ARPES and density-functional based augmented spherical wave calculations have revealed the semimetallic nature of WTe${}_{2}$~\cite{Augustin00PRB}. However, no details close to the Fermi level were clearly resolved. More recent, high resolution ARPES data has revealed one pair of electron pockets and one pair of hole pockets of similar size, supporting the electron-hole carrier compensation theory as the primary origin of the extremely large magnetoresistance~\cite{Pletikosic14PRL}. By varying the incident photon energies in the 40--70~eV range, the ${k}_{z}$ dispersion of the states was mapped out with some bands showing low dispersion and some showing variations in intensity, but no solid conclusion can be drawn from these data~\cite{Pletikosic14PRL}. Another study reported presence of nine Fermi pockets. However, no significant photon energy dependence along the out of plane direction was observed~\cite{Jiang2015Signature}. On the other hand, magnetoresistance measurements with varying magnetic field applied at an angle with respect to the $c$ axis of the sample have led to the conclusion of three-dimensional electronic structure in WTe${}_{2}$~\cite{Thoutam2015Temperature}. Furthermore, the results from quantum oscillations--another technique to probe the Fermi surface structure--have come to similar conclusions. Angle resolved quantum oscillation measurements implied strong three-dimensionality of the band structure in this material~\cite{Zhu15PRL}. The analysis of quantum oscillations is a powerful technique to measure the extrema of the Fermi surface topology. However, one should be cautious when assigning oscillation frequencies to particular FS pockets, e.g. in measurements performed under applied pressure, one group assigned the peaks that survived at high pressure to a particular pair of electron and hole pockets~\cite{Cai15PRL}, however, the Hall effect measurements from another group found that only electron carriers were present under high pressure~\cite{Kang15NatComm}, a result that is consistent with our report of temperature-induced Lifshitz transition in WTe${}_{2}$~\cite{Wu15PRL}. Thus, in order to demonstrate the three-dimensionality of the electronic structure in WTe${}_{2}$, ultrahigh resolution ARPES measurements with fine tunable incident photon energies have real experimental advantage. 

Here, we use temperature- and field-dependent resistivity measurements and ultrahigh resolution, tunable VUV laser-based ARPES to probe the three-dimensionality of the bulk electronic structure in WTe${}_{2}$. In the SdH oscillations, a low frequency peak was observed and can be explained by our photon energy dependent ARPES measurements. With the ability of fine tuning of the incident photon energy from 5.77 to 6.7~eV, we have determined Fermi surface and band structure with very high precision. At the incident photon energy of 6.7~eV, we can detect a bulk FS that consists of two pairs of electron pockets and two pairs of hole pockets, while the top of another band is located just below the Fermi level at the $\Gamma$ point. When decreasing the incident photon energy to 6.36~eV, another pair of tiny electron pockets is detected close to the $\Gamma$ point, which corresponds to the so far unaccounted for, low oscillation frequency observed in the quantum oscillation measurements~\cite{Cai15PRL, Wu15PRL}. Further decreasing the incident photon energy, we observe the disappearance of the tiny electron pockets, thus the bulk FS has only two pairs of electron and hole pockets left for this range of ${k}_{z}$ momenta. Detailed band dispersion along several cuts for different incident photon energies are presented, demonstrating the strong three-dimensionality of the bulk electronic structure in WTe${}_{2}$. These results are consistent with the band structure calculations and quantum oscillations~\cite{Ali14Nat, Cai15PRL, Wu15PRL}. Our photon energy dependent ARPES measurements have solved the mystery of the low frequency peak reported by several quantum oscillation measurements~\cite{Cai15PRL, Wu15PRL}.

\section{Experimental details}
Single crystals of WTe${}_{2}$ were grown by solution method~\cite{Canfield92PMPB, Canfield2016Use}, following the procedure describe in Ref.~\onlinecite{Wu15PRL}. The resulting crystals were blade or ribbon like in morphology with typical dimensions of 3 $\times$ 0.5 $\times$ 0.01~mm with the crystallographic $c$ axis being perpendicular to the crystal surface; the crystals are readily cleaved along this crystal surface.

Magnetic field dependent electrical transport measurements were carried out in a Quantum Design Physical Property Measurement System (PPMS) for $H \le 14$~T. Samples for standard four-probe resistivity measurement were prepared by attaching four Pt wires using Epotek-H20E silver epoxy. The field was applied parallel to the crystallographic $c$ axis, and the current was along the crystallographic $a$ axis. Magnetoresistance was measured at 1.8, 2.5, 4, 6, 8, and 10~K.

Samples used for ARPES measurements were cleaved \textit{in situ} at 40~K under ultrahigh vacuum (UHV). The data were acquired using a tunable VUV laser ARPES system, that consists of a Scienta R8000 electron analyzer, picosecond Ti:Sapphire oscillator and fourth harmonic generator~\cite{Jiang14RSI}. Data were collected with a tunable photon energies from 5.3 to 6.7~eV. Momentum and energy resolutions were set at $\sim$ 0.005~\AA${}^{-1}$ and 2~meV. The size of the photon beam on the sample was $\sim$30~$\mu$m.

\section{Experimental results} 

\begin{figure}
	\includegraphics[width = 3in]{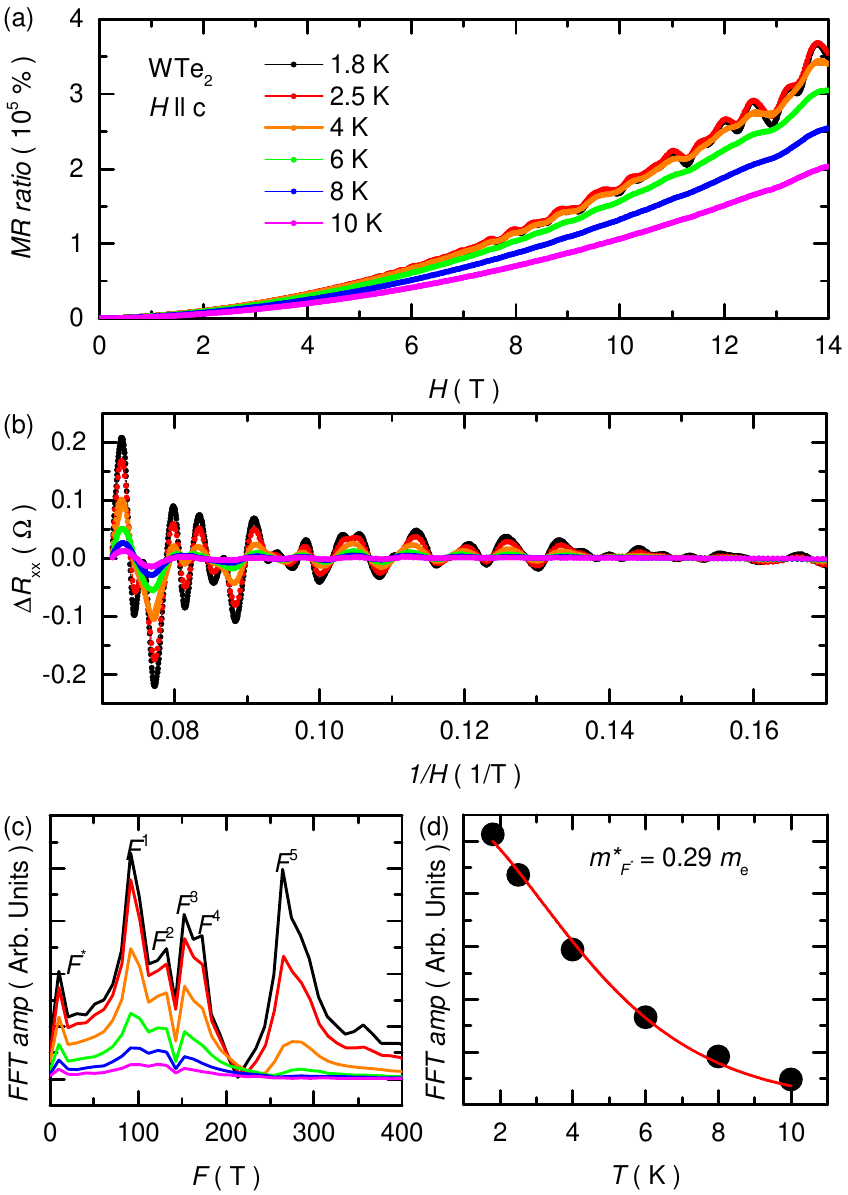}%
	\caption{(color online) Quantum oscillation analysis on WTe$_{2}$. 
	(a) Magnetoresistance measured at $T=$1.8, 2.5, 4, 6, 8, and 10~K.
	(b) Shubnikov-de Haas oscillation after subtracting the background. 
	(c) Fast Fourier transform (FFT) analysis of quantum oscillation. 
	(d) Temperature dependence of the oscillation amplitude as a function of temperatures for the peak, $F^{*}$. The closed circles are the data and solid line is the fitted line of Lifshitz-Kosevich formula. 
	\label{fig:Fig1}}
\end{figure}

\begin{figure*}[bt]
	\includegraphics[width = 6in]{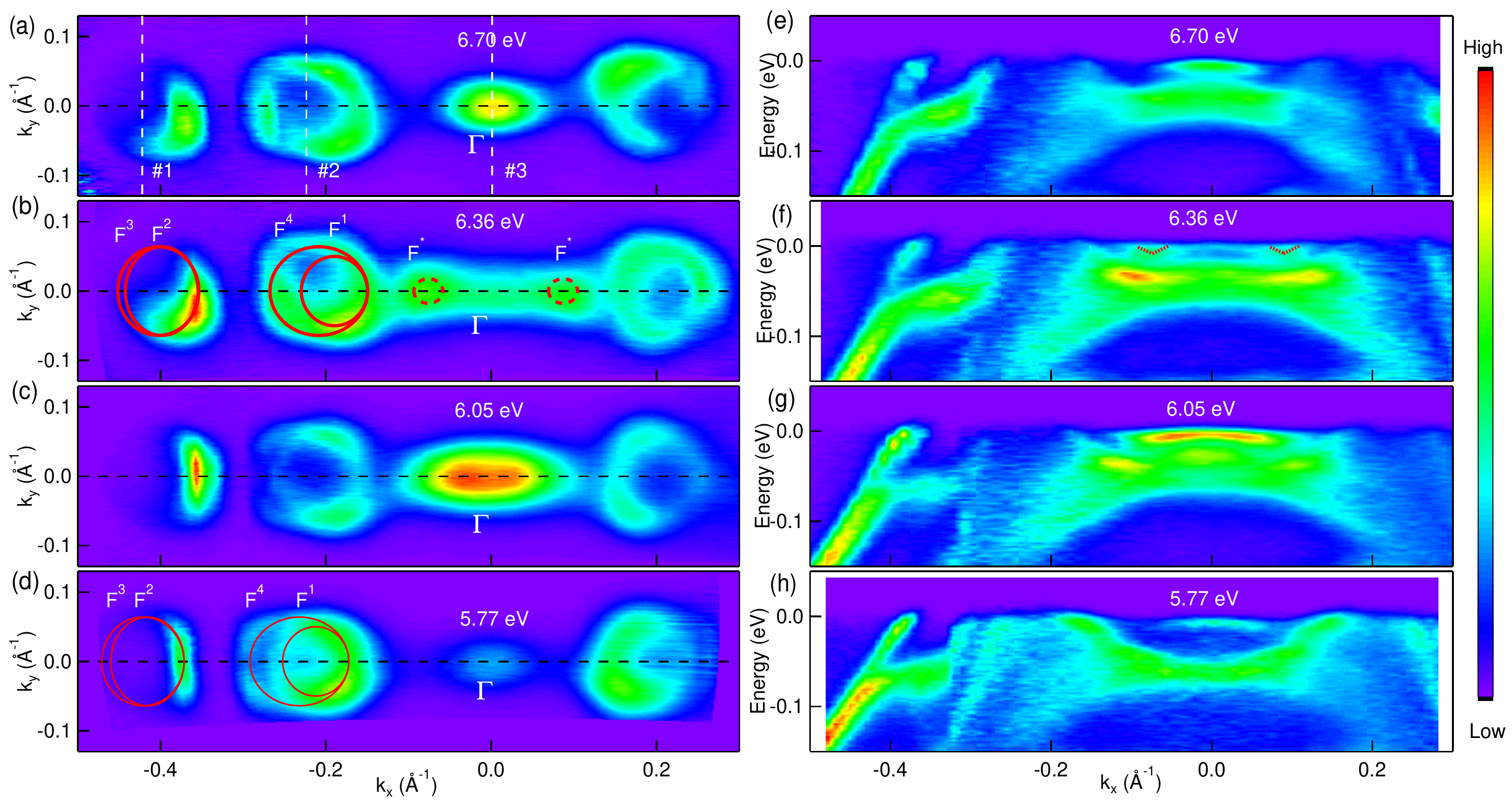}%
	\caption{Fermi surface plots and band dispersion measured at different photon energies.
	(a)-(d) Fermi surface plots measured at photon energies of 6.70, 6.36, 6.05, and 5.77 eV, respectively. The red solid circles (from left to right) correspond to the quantum oscillation frequencies of ${F}^{3}$, ${F}^{2}$, ${F}^{4}$, and ${F}^{1}$, respectively. The red dashed circles correspond to the the quantum oscillation frequency of ${F}^{*}$.
	(e)-(h) Band dispersion along the black dashed lines in (a)-(d), respectively. The red dashed lines are guide to the eye.
	\label{fig:Fig2}}
\end{figure*}

\begin{figure*}[tb]
	\includegraphics[width = 6.5in]{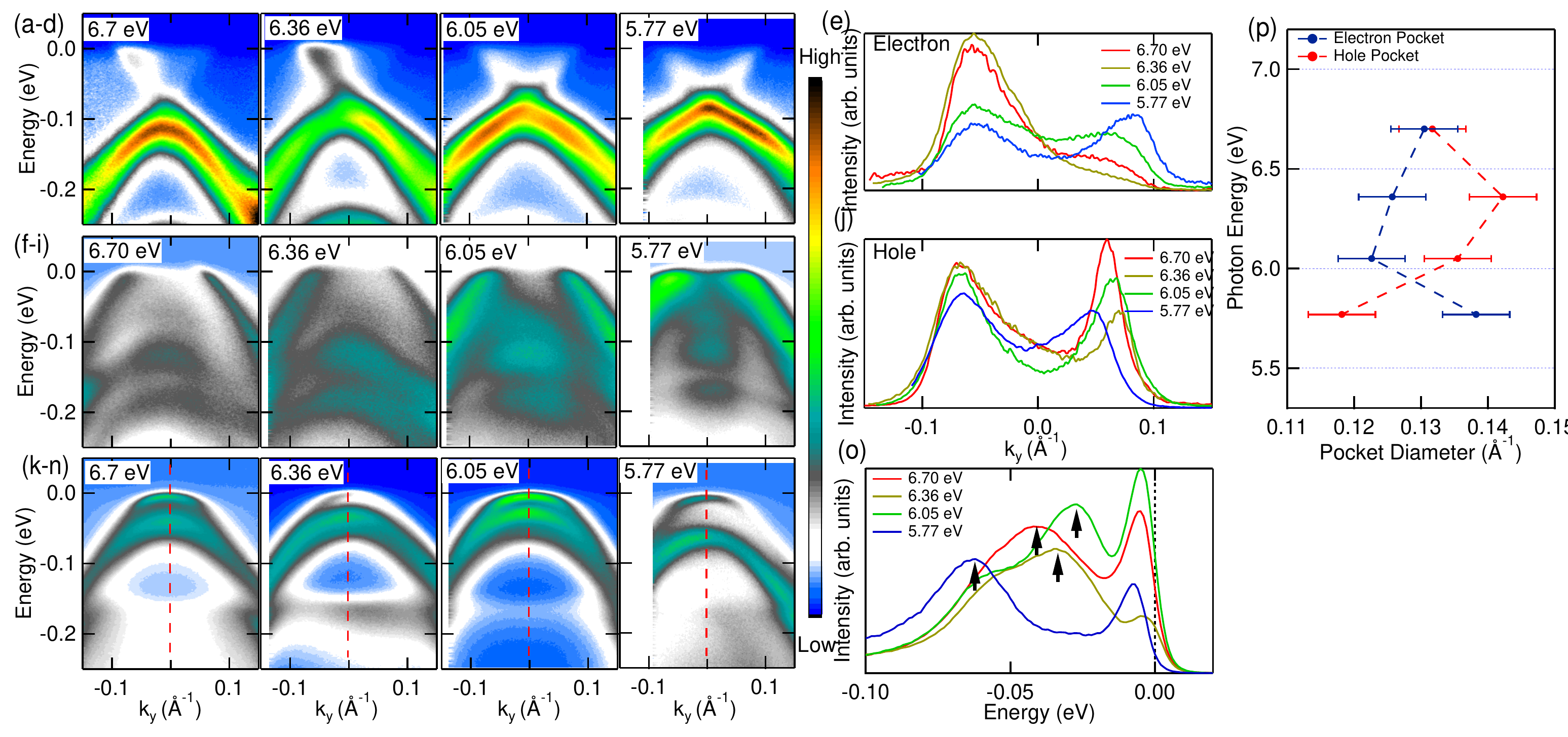}%
	\caption{Band dispersion, momentum dispersion curves, and energy dispersion curves measured at different photon energies.
	(a)-(d) Band dispersion along the cut \#1 in Fig.~\ref{fig:Fig2}(a) measured at photon energies of 6.7, 6.36, 6.05, and 5.77 eV, respectively. 
	(e) Momentum dispersion curves at the E${}_{F}$ of (a)-(d).
	(f)-(i) Band dispersion along the cut \#2 in Fig.~\ref{fig:Fig2}(a) measured at photon energies of 6.7, 6.36, 6.05, and 5.77 eV, respectively. 
	(j) Momentum dispersion curves at the E${}_{F}$ of (f)-(i).
	(k)-(n) Band dispersion along the cut \#3 in Fig.~\ref{fig:Fig2}(a) measured at photon energies of 6.7, 6.36, 6.05, and 5.77 eV, respectively. 
	(o) Energy dispersion curves along the red dashed lines in (k)-(n). Black arrows point to the locations of the lower hole bands in (k)-(n).
	(p) Diameters of the electron and hole pockets measured at different photon energies extracted from the momentum dispersion curves in (e) and (j).
	\label{fig:Fig3}}
\end{figure*}

Magnetoresistance (MR) shows parabolic behavior without any saturation at high field as shown in Fig.~\ref{fig:Fig1}(a). In order to analyze the quantum oscillation spectra, we subtracted the background using a second order polynomial function to fit the background MR in the range of 6~$\le H \leq$~14~T for all temperatures. The oscillations show periodic behavior in 1/$H$ as shown in Fig.~\ref{fig:Fig1}(b). 

The frequencies of the oscillation were obtained by FFT analysis shown in Fig.~\ref{fig:Fig1}(c). Five frequencies, including ${F}^{1} =92~\textrm{T}$, ${F}^{2} = 132~\textrm{T}$, ${F}^{3} = 152~\textrm{T}$, ${F}^{4} = 172~\textrm{T}$ and ${F}^{5} = 264~\textrm{T}$, are similar to published results~\cite{Cai15PRL, Zhu15PRL, Wu15PRL}. Interestingly, a new low frequency peak, ${F}^{*} = 10~\textrm{T}$, is also clearly observed in our data. The amplitude of the ${F}^{*}$ decreases with increasing temperature as shown in Fig.~\ref{fig:Fig1}(d). Note that for the sake of consistency the exactly same data acquisition and processing protocol was followed for all temperatures. The observed decrease of the oscillation amplitude is due to the temperature induced scattering of electrons, described by the Lifshitz-Kosevich formula~\cite{Shoenberg1984Magnetic}:

\begin{equation}
{ A }_{ t }\propto { B }^{ 1/2 }{ \left| \frac { { \partial  }^{ 2 }S_{ t } }{ \partial { k }_{ H }^{ 2 } }  \right|  }^{ -1/2 }{ R }_{ T }{ R }_{ D }{ R }_{ S }
\end{equation} 

\noindent where the factor ${R}_{T}$ is related to thermal damping, ${R}_{D}$ is related to impurities and ${R}_{S}$ is related to spin Zeeman splitting and superposition of spin-up and spin-down oscillations. The thermal damping part ${R}_{T}$ is defined as 

\begin{equation}
{ R }_{ T }=\frac { \alpha { m }^{ * }T/B }{ sinh(\alpha { m }^{ * }T/B) } 
\end{equation} 

\noindent where $\alpha = 2 {\pi}^{2} c {k}_{B} / e \hbar$. Using this equation, we calculated the effective mass of the carriers linked to the oscillation frequency ${F}^{*}$, ${m}^{*}_{{F}^{*}} = 0.29 \pm 0.01~{m}_{e}$.

To match this small frequency observed in quantum oscillation to a specific Fermi surface, we carried out photon energy dependent ARPES measurements. By varying the incident photon energies, we are able to map out the band dispersion along the out of plane, ${k}_{z}$ direction~\cite{Damascelli03RMP,Campuzano04PS}. Synchrotron radiation based ARPES systems are often used for ${k}_{z}$ dispersion mapping due to the large tunable range of the incident photon energies. However, tuning photon energies with usually utilized coarse steps $\ge$~1~eV can result in some important details being missed along very key ${k}_{z}$ direction~\cite{Pletikosic14PRL, Jiang2015Signature}. By using tunable VUV laser ARPES with very fine energy steps, we mapped out the ${k}_{z}$ dispersion of WTe${}_{2}$ in great detail. Fig.~\ref{fig:Fig2} shows the FS and band dispersion measured using incident photon energies of 6.70, 6.36, 6.05, and 5.77~eV, as indicated at the top center of each plot. In Figs.~\ref{fig:Fig2}(a)-(d), we can see that the FS of WTe${}_{2}$ measured using different photon energies look similar, with two pairs of electron pockets and two pairs of hole pockets in the first BZ. However, a significant variation between the data sets is also observed. The hole band at the $\Gamma$ point has different intensities and curvatures, although none of them crosses the Fermi level as shown in Figs.~\ref{fig:Fig2}(e)-(h). Furthermore, the FS close to the $\Gamma$ point in Fig.~\ref{fig:Fig2}(b) shows a dumb-bell like structure, whereas the other three FSs show only a single hole band at the $\Gamma$ point. The band dispersion along the black dashed line in Fig.~\ref{fig:Fig2}(b) is shown in panel (f). On either side of the $\Gamma$ point, a tiny electron pocket is visible [marked by the red dashed lines in panel (f)], which is different from the band dispersion observed using other photon energies. 

To directly compare the quantum oscillation results with the ARPES measurements, we have plotted the extremal orbits with the areas determined from quantum oscillation measurements on top of the Fermi surface plot as shown in Fig.~\ref{fig:Fig2}(b). The corresponding extremal areas of the Fermi surface were calculated using the Onsager relation, ${F}^{i} = \frac{\hbar c} {2 \pi e} {S}^{i}$~[Ref.~\onlinecite{Shoenberg1984Magnetic}] with ${S}_{{F}^{1}} = 0.00874~{{\textrm{\AA}}^{-2}}$, ${S}_{{F}^{2}} = 0.01262~{{\textrm{\AA}}^{-2}}$, ${S}_{{F}^{3}} = 0.01456~{{\textrm{\AA}}^{-2}}$, ${S}_{{F}^{4}} = 0.0165~{{\textrm{\AA}}^{-2}}$, and ${S}_{{F}^{*}} = 0.000971~{{\textrm{\AA}}^{-2}}$. For simplicity, we assume that the areas obtained from quantum oscillations are from simple circle/ellipse orbit extrema and the corresponding shapes are plotted in Fig.~\ref{fig:Fig2}(b). We can clearly see a good match between the quantum oscillation results and ARPES measurements of the Fermi surface. Furthermore, the electronic structure calculations that take into account the spin-orbit coupling in WTe${}_{2}$ bring extra tiny electron pockets close to the $\Gamma$ point~\cite{Ali14Nat}. Thus, our ARPES results have solved the mystery of this unaccounted for, low frequency quantum oscillation peaks. We should note that as the electronic structure of WTe${}_{2}$ is very sensitive to pressure/strain~\cite{Kang15NatComm, Pan15NatComm, Soluyanov2015Type, Wu2016Observation}, it is possible that these tiny electron pockets might be suppressed in some of the quantum oscillation or ARPES measurements~\cite{Zhu2015Quantum, Wu2016Observation}

Figure~\ref{fig:Fig3} shows the detailed band dispersion measured using various photon energies. Panels (a)--(d), (f)--(i), and (k)--(n) present the band dispersions measured using photon energy of 6.7, 6.36, 6.05 and 5.77~eV, respectively, and correspond to the cuts as marked \#1, 2, and 3 in Fig.~\ref{fig:Fig2}(a). In panels (a)--(d), only minor intensity differences can be seen between the four measurements. At the photon energy of 6.05 and 5.77 eV, the electron pockets are clear and symmetric. On the other hand, the electron pockets measured at the photon energy of 6.7 and 6.36 eV are not symmetric in intensity, probably due to the matrix elements effect. To quantify the electron pocket sizes in panels (a)--(d), we have plotted the momentum distribution curves (MDCs) at the Fermi level ${E}_{F}$ in Fig.~\ref{fig:Fig3}(e). The peak locations of the MDCs show clear differences across these four photon energies (peak locations of the left branches are aligned for easy comparison). Panels (f)--(i) present the band dispersion from cut \#2, which clearly shows that the hole pocket measured with 5.77 eV photons is significantly smaller than the other three. Panel (j) shows the MDCs at the ${E}_{F}$ from panels (f)--(i), illustrating different hole pocket sizes [also left aligned as in (e)]. Panels (k)--(n) show the photon energy dependence of the hole bands at the $\Gamma$ point [cut \#3 in Fig.~\ref{fig:Fig2}(a)]. Two hole bands can be clearly seen at the $\Gamma$ point with different separations between them for different photon energies. Panel (o) shows the energy distribution curves (EDCs) from panels (k)--(n), where the black arrows point to the peak locations in the lower hole bands. The upper hole bands sit at roughly the same binding energy for these photon energies, but none of them crosses the Fermi level. On the other hand, the distance between the upper and lower hole bands is very different across these photon energies, with 5.77 eV showing the maximum separation. By fitting two Lorentzian functions to the MDCs in panels (e) and (j), we calculate the electron/hole pocket sizes and summarize the results in panel (p). With the decreasing incident photon energy, the size of the electron pocket decreases and then increases. On the other hand, the size of the hole pocket increases and then decreases. This trend (strong three-dimensionality along ${k}_{z}$ direction) is consistent with the band structure calculations shown in Ref.~\onlinecite{Ali14Nat}, where the hole pockets have a concave shape and the electron pockets have a convex shape along ${k}_{z}$ direction toward the center of the zone.

\section{Conclusion}
In conclusion, we used temperature- and field-dependent resistivity measurements in tandem with ultrahigh resolution laser ARPES to investigate the detailed electronic structure of WTe${}_{2}$. The photon energy dependence measurements with relatively fine energy steps have revealed the three-dimensional character of the electron and hole pockets along the $\Gamma-Z$ direction. With the increase of the incident photon energy from 5.77 to 6.70~eV (i.e., probing along ${k}_{z}$ direction), we have observed that the hole pocket expands and then shrinks, while the electron pocket displays opposite behavior. Strong photon energy dependence is also observed in the hole bands at the $\Gamma$ point. Furthermore, at the photon energy of 6.36~eV we have revealed a pair of tiny electron pockets sitting at the opposite side of the $\Gamma$ point, providing strong support for the low quantum oscillation frequency peak that was not accounted for in the previous studies~\cite{Wu15PRL, Cai15PRL}. 

\begin{acknowledgements}
Research was supported by the US Department of Energy, Office of Basic Energy Sciences, Devision of Materials Science and Engineering. Ames Laboratory is operated for the US Department of Energy by the Iowa State University under Contract No. DE-AC02-07CH11358. Y.W. (Analysis of ARPES data) was supported by Ames Laboratory’s Laboratory-Directed Research and Development (LDRD) funding. N.H.J. was supported by the Gordon and Betty Moore Foundation EPiQS Initiative (Grant No. GBMF4411). L.H. was supported by CEM, a NSF MRSEC, under Grant No. DMR-1420451. 
\end{acknowledgements}

Raw data for this manuscript is available at \url{http://lib.dr.iastate.edu/ameslab_datasets/}.

\bibliography{WTe2_3D}

\end{document}